# Transfer entropy dependent on distance among agents in quantifying leader-follower relationships

Udoy S. Basak[1,2], Sulimon Sattari[3], Md. Motaleb Hossain[3,4], Kazuki Horikawa[5] and Tamiki Komatsuzaki[1,3,6,7]

[1]Graduate School of Life Science, Transdisciplinary Life Science Course, Hokkaido University, Sapporo 060-0812, Japan
[2]Pabna University of Science and Technology, Pabna 6600, Bangladesh
[3]Research Center of Mathematics for Social Creativity, Research Institute for Electronic Science, Hokkaido University, Kita 20, Nishi 10, Kita-ku, Sapporo 001-0020, Japan
[4]University of Dhaka, Dhaka 1000, Bangladesh
[5]Department of Optical Imaging, The Institute of Biomedical Sciences, Tokushima University Graduate School, 3-18-15 Kuramoto-cho, Tokushima City, Tokushima 770-8503, Japan
[6]Institute for Chemical Reaction Design and Discovery (WPI-ICReDD), Hokkaido University, Kita 21 Nishi 10, Kita-ku, Sapporo, Hokkaido 001-0021, Japan
[7]Graduate School of Chemical Sciences and Engineering Materials Chemistry and Engineering Course, Hokkaido University, Kita 13 Nishi 8, Kita-ku, Sapporo 060-0812, Japan

**Synchronized movement of (both unicellular and multicellular) systems can be observed almost everywhere. Understanding of how organisms are regulated to synchronized behavior is one of the challenging issues in the field of collective motion. It is hypothesized that one or a few agents in a group regulate(s) the dynamics of the whole collective, known as leader(s). The identification of the leader (influential) agent(s) is very crucial. This article reviews different mathematical models that represent different types of leadership. We focus on the improvement of the leader-follower classification problem. It was found using a simulation model that the use of interaction domain information significantly improves the leader-follower classification ability using both linear schemes and information-theoretic schemes for quantifying influence. This article also reviews different schemes that can be used to identify the interaction domain using the motion data of agents.**

### Introduction:

Collectives are ubiquitous in many biological systems from unicellular organisms, e.g., the "social" amoeba dictyostelium discoideum [1,2], cancer development [3], and wound healing [4], to multicellular systems such as flocks of birds [5,6] and schools of fish [7]. How can one infer the underlying mechanism of regulation among individuals? Trajectories of agents within their group movement are one of the devices to shed light on unveiling the rule(s). Several simulation models have been proposed for shedding light on unveiling rules which regulate collective motility [8,9]. These models have facilitated the analysis of collective motion [10]. The Vicsek model (VM) [11] is one of the most well-studied models, based on a simple rule where the neighboring agents tend to align to their direction to the average over those positioned within an interaction domain. The VM was found to be suitable in studying the underlying origin of dynamics of collectives such as symmetry breaking [12,13] and phase transition [14,15]. More recently, the Vicsek model has been modified to include leadership in order to characterize the influence of leaders [9,16,17].

It is natural to imagine that two agents cannot communicate over an infinite distance. Two agents can influence only each other when they are located within each other's interaction domain. In the VM, the interaction domain is a circle of radius $R$. Knowledge of interaction domain plays a vital rule in the classification of leaders and followers.

It has been shown both theoretically and experimentally that the collective movement of agents is guided by dominant individuals, termed as 'leaders', which control the movement of the whole [3,4,16,18]. These special agents employ asymmetric influence on the other group members, termed as 'followers'. There, causal inference is an essential aspect of determining leadership and its role in collective systems. The leader-follower relationship between agents can be observed in the wound-healing procedure [4], cancer growth [3], and MDCK epithelial cell migration [18,19]. For example, for MDCK epithelial cells the removal of leader cell(s) causes a disturbance in the cell migration [18]. Identifying leader and follower agents is, however, a complex task. Sometimes the relative positions of agents within a moving group can be indicative of the leaders of the group. For example, in the migration of MDCK epithelial cells, leader cells are positioned at the tips of finger-like structures [18].

Corresponding author: Tamiki Komatsuzaki, Research Center of Mathematics for Social Creativity, Research Institute for Electronic Science, Hokkaido University, Kita 20 Nishi 10, Kita ku, Sapporo 001-0020, Japan.
e-mail: tamiki@es.hokudai.ac.jp

***Significance***

Synchronized movement of (both unicellular and multicellular) systems can be observed almost everywhere. It is hypothesized that one or a few agents in a group regulate(s) the dynamics of the whole collective, known as leader(s). Leader(s) may provide vital information about the system. Hence the identification of the leader (influential) agent(s) is very crucial. In this article, we focus on the improvement of the leader-follower classification problem using only the motion data of agents.

In the case of wound healing, the procedure starts with the formation of mounds and one or rarely two leader cells positioned at the top of the mound pull neighboring follower cells forward into the wound [4]. Also, in cancer growth, the leader cells were found at the front of the 'invasive strands' [3]. For a fish shoal, the front fish has a strong influence on the movement of the shoal [20].

But it is not necessary that the leader should always be at the front—it may change its position over time. Also, in many cases it is not possible to identify the top or front cells, e.g., in dictyostelium discoideum amoeba migration where the positions of cells may have no meaning in terms of leader-follower identification.

For such cases, experimental data e.g., the ensemble of trajectories of agents, can be used to infer distinctive influences in their interactions and in classifying leader and follower agents. Various types of linear measures such as cross-correlation [9,21] and Granger causality [22,23] have been used on time-lapse motion data to quantify the direction of influence between pairs of agents. However, interaction between two agents in nature is generally non-linear, implying that linear measures are not necessarily optimal in capturing underlying interactions. As an alternative approach, different information-theoretic schemes such as mutual information [24], time-delayed mutual information [25], transfer entropy (TE) [26], and causation entropy [27] have been intensively studied, which are all based on probability distributions and are free from assuming linear interactions.

This article reviews two central issues in the field of leader-follower classification:

i) Solely from measurements, how can one infer causal relationships between leaders and followers?

ii) How can one infer the interaction domain solely from the motion data of the agents?

This paper is organized as follows: in 'Measuring causality' section we first overview the possible causal relationship between a leader agent and a follower agent. Concurrently, we address the different schemes in quantifying causality between two processes. Then in the 'Models' section, we present a modified version of the Vicsek model to elucidate the underlying leader-follower relationship among a group of collectively moving agents. Also, different types of leadership are addressed. We then present different types of classifier to identify the leader agent(s) in a group. Later, we review some existing results in classifying leader and follower agents, and also in identifying 'interaction domain'. Finally, we provide some future directions in the 'Conclusion and perspectives' section.

**Measuring Causality:**

In the literature, different definitions of leadership can be found. In a group of moving agents, the agent moving in the front position during the movement is often considered as the leader of the group [28,29]. The agent that departs 'first' or has the ability to guide the whole group in adverse conditions (e.g., famine, presence of predators) is also considered as a leader [30,31]. In this article, we follow the definition of leadership by Krause et al. [20] 'as the initiation of new directions of locomotion by one or more individuals which are then readily followed by other group members'. Based on this definition of leadership, a causal relationship between a leader and a follower can be assumed and may be detectable based on trajectory data. Hence by identifying the causal direction in a pair of agents, the leader and follower relationship can be inferred within the pair. The agent which, on average, leads other group members can be identified as the group leader [16].

**Linear schemes:**

Linear correlation is a measure which indicates how two or more variables are linearly related. Suppose that $X = \{\ldots, x_{t-1}, x_t, x_{t+1}, \ldots\}$ and $Y = \{\ldots, y_{t-1}, y_t, y_{t+1}, \ldots\}$ are two random variables. Then the correlation coefficient has the following form [32]:

$$r = \frac{\sum_t[(x_t - \bar{x})(y_t - \bar{y})]}{\sqrt{\sum_t(x_t - \bar{x})^2}\sqrt{\sum_t(y_t - \bar{y})^2}},$$

where $\bar{x}$ and $\bar{y}$ denote the time averages of $\{x_t\}$ and $\{y_t\}$, respectively.

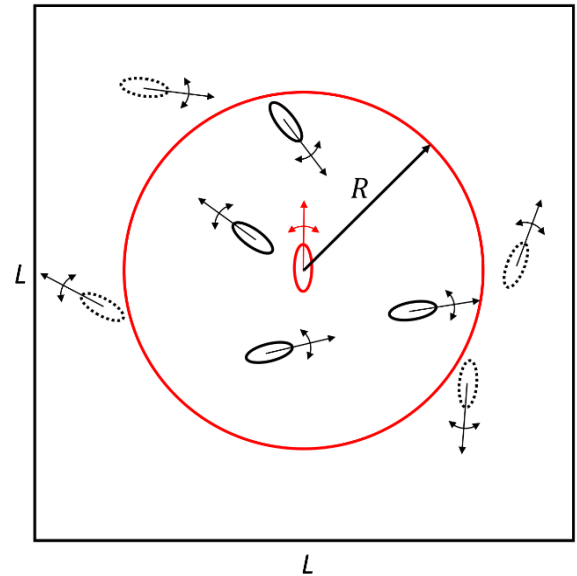

**Figure 1** The ovals (solid and dotted) denote the moving agents within a square box of length $L$, and the red circle of radius $R$ represents the interaction domain of the red agent (at the center of the circle). According to the protocol used in the VM [11], the motion of the red agent is affected by the solid-black agents as they are located within the interaction domain of the red agent. But the black dotted ovals are located outside the interaction domain, hence the motion of the red agent is free from the influence of those dotted-black agents.

The value of correlation coefficient $r$ lies between -1 and +1. $r$ takes the value of 1 (-1) if the two variables are completely correlated positively (negatively), and if the variables are completely independent the coefficient $r$ is zero [32]. The correlation coefficient does not shed any light on the underlying associations behind the relationships between variables (i.e. causal relationships), rather it just tells if linear correlation exists between them.

Cross-correlation is a measure of similarity between two processes as a function of the displacement (delay) of one relative to another. Cross-correlation between two random variables $X$ and $Y$ as a function of time-delay $\tau$ has the following form:

$$C_{XY}(\tau) = \frac{\sum_t[(x_t - \bar{x})(y_{t+\tau} - \bar{y})]}{\sqrt{\sum_t(x_t - \bar{x})^2}\sqrt{\sum_t(y_{t+\tau} - \bar{y})^2}}. \quad (1)$$

It has successfully been used to identify if some individuals have a strong influence on other agents [16, 20]. For instance, if there exists a positive delay time $\tau$ that maximizes $C_{XY}(\tau)$, one may interpret that at the level of linearity $X$ has some influence over $Y$.

The primary step to finding the leader is to measure pairwise causality within the group. Measuring causality has been one of the most intriguing subjects since it cannot always be proven. Wiener [33] first proposed a quantifiable definition of causality. It was proposed that for one variable to be "causal" to another, the information about the first variable should improve the predictability of the second variable [34,35]. The first variable is known as the 'cause' and the second one is called the 'effect'. However, no practical implementation scheme was presented. Later Granger [22] proposed a scheme applicable to real data. Suppose that we want to identify whether there is a causal relationship between variables $X$ and $Y$. In order for a causal relationship to be inferred, knowledge of one variable should improve one's ability to predict the other one. Suppose we want to predict $x_{t+1}$. Suppose that, to predict $x_{t+1}$, one can use either only the past terms of $X$ or those of both $X$ and $Y$. If the second prediction is more accurate compared to the first one, then one may conclude that the past of $Y$ contains information that is helpful to predict $x_{t+1}$ which is not contained in the past of $X$. In this case, $Y$ is said to be a G-cause (Granger cause) of $X$ [22].

Granger causality is normally tested based on linear regression models as follows. The prediction of $x_t$ can be made based on the following linear autoregression on its own history:

$$x_t = \sum_{i=1}^{\infty} a_i x_{t-i} + \delta_t,$$

where $\{a_i\}$ represent the autoregression coefficients and $\{\delta_t\}$ characterize the corresponding prediction errors [36].

Moreover, one can use the history of both $X$ and $Y$ to predict $x_t$ using linear regression as follows:

$$x_t = \sum_{i=1}^{\infty} b_i x_{t-i} + \sum_{j=1}^{\infty} c_j y_{t-j} + \bar{\delta}_t,$$

where $\{b_i\}$ and $\{c_j\}$ are the joint linear regression coefficients, and $\{\bar{\delta}_t\}$ are the corresponding prediction errors. If $X$ is causally driven by $Y$, i.e., $Y$ is the cause and $X$ is the effect, then the prediction of $x_t$ is expected to be improved by incorporating the history $Y$ along with the history of $X$, compared to using the history of $X$ alone. Consequently, the variance of the prediction error $\bar{\delta}_t$ is expected to be smaller than that of $\delta_t$. Hence the Granger causal value from $Y$ to $X$ has the following form [36]:

$$G_{Y \to X} = \log \frac{\text{var}(\delta_t)}{\text{var}(\bar{\delta}_t)}.$$

In the case where $Y$ has no causal influence on $X$, the prediction of $x_t$ is not improved by the incorporation of the history of $Y$, which means that the prediction errors would have same variances, i.e., $\text{var}(\delta_t) \simeq \text{var}(\bar{\delta}_t)$ that results in zero Granger causality from $Y$ to $X$, i.e., $G_{Y \to X} \simeq 0$.

When there is causal relationship between two variables, there must exist a certain time lag (delay) between the cause and the effect [37]. The Wiener-Granger framework of prediction-based causality is equivalent to looking for dependencies between the variables at a certain time delay [38].

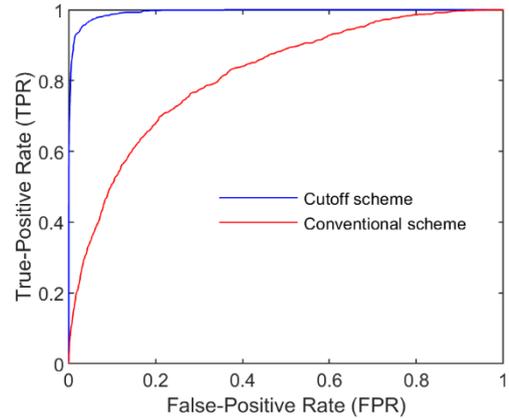

**Figure 2** ROC curves using the modified VM [9] for $R = 2$ at $\eta_0 = 4\pi/3$ using two different schemes. The ROC using the 'cutoff scheme' lies above that of 'without cutoff scheme'. It suggests that the cutoff scheme possesses better classification ability compared to the without cutoff (conventional) scheme.

Both Granger causality and cross-correlation are based on a linear relationship between the two variables. Dynamical systems can be highly nonlinear, in general, suggesting that the application of Granger causality and cross-correlation is limited to some classes of systems [36].

**Information-theoretic schemes:**

Information-theoretic measures are based on probability distributions of stochastic random variables; hence these measures are model-free and expected to be capable of capturing nonlinear interactions between systems [39].

Suppose the probability mass functions of $X$ and $Y$ are $p(x_t) = \Pr(X = x_t)$ and $p(y_t) = \Pr(Y = y_t)$, respectively. The fundamental measure of information in information theory is *Shannon entropy* or simply *entropy*

$H(X)$, which is defined by [24]

$$H(X) = -\sum_{x_t} p(x_t) \log_2 p(x_t).$$

The unit of entropy depends on the base of the logarithm. In this article we use the logarithm of base 2 so that the unit of information is in *bits*. $H(X)$ measures the uncertainty associated with the variable $X$. $H(X)$ is maximum whenever the outcomes of the random variable $X$ are all equally probable. On the other hand, if only one outcome is more likely to happen and all the rest are not, $H(X)$ is zero.

The joint entropy of $X$ and $Y$, denoted by $H(X,Y)$, is defined by [24]

$$H(X,Y) = -\sum_{x_t}\sum_{y_t} p(x_t, y_t) \log_2 p(x_t, y_t),$$

where $p(x_t, y_t)$ represents the joint probability distribution of $X$ and $Y$. Joint entropy $H(X,Y)$ represents the amount of uncertainty associated with both processes $X$ and $Y$.

Conditional entropy of $X$ given $Y$, denoted by $H(X|Y)$, measures the remaining uncertainty about the random variable $X$ when the outcomes of the other variable $Y$ are known [24]:

$$H(X|Y) = -\sum_{x_t}\sum_{y_t} p(x_t, y_t) \log_2 p(x_t|y_t),$$

where $p(x_t|y_t)$ represents the conditional probability of $X$ given $Y$. If the random variable $Y$ has no information about $X$, the knowledge of $Y$ does not reduce the uncertainty of $X$. In that case, $H(X|Y) = H(X)$.

The random variables $X$ and $Y$ can be independent of, or dependent on each other. In the case of dependency, one random variable contains information about another. To quantify the amount of information that $X$ has about $Y$ (or vice versa) one can compute mutual information [24]:

$$I(X;Y) = \sum_{x_t}\sum_{y_t} p(x_t, y_t) \log_2 \frac{p(x_t, y_t)}{p(x_t)p(y_t)}. \quad (2)$$

Equation (2) means that mutual information is symmetric, i.e., $I(X;Y) = I(Y;X)$. For the case of causality, the present of a variable (effect) depends on the past of another variable (cause) and the causal direction (direction of information flow) from the cause to the effect is inferred. One may obtain an asymmetric measure named time-delayed mutual information by introducing a time-lag parameter $\tau$ in any of the variables $X$ and $Y$ as follows [40,41]:

$$I(X(t); Y(t+\tau)) = \sum_{x_t, y_{t+\tau}} p(x_t, y_{t+\tau}) \log_2 \frac{p(x_t, y_{t+\tau})}{p(x_t)p(y_{t+\tau})}, \quad (2a)$$

and

$$I(X(t+\tau); Y(t)) = \sum_{x_{t+\tau}, y_t} p(x_{t+\tau}, y_t) \log_2 \frac{p(x_{t+\tau}, y_t)}{p(x_{t+\tau})p(y_t)}. \quad (2b)$$

A non-zero amplitude of $I(X(t); Y(t+\tau))$ as a function of delay-time $\tau^*$ indicates the presence of interaction between the processes $X$ and $Y$, and the sign of $\tau$ can be used to infer the direction of influence [36]. If there exists a positive $\tau^*$ for which time-delayed mutual information has a peak as a function of $\tau$ (Eq. (2a)) one can deduce that $X$ shares a maximum amount of information with the future state of $Y$. In other words, one may say that $X$ drives $Y$. Conversely, a negative $\tau$ indicates that $X$ shares a maximum amount of information with the past of $Y$, and hence $X$ is driven by $Y$ [36].

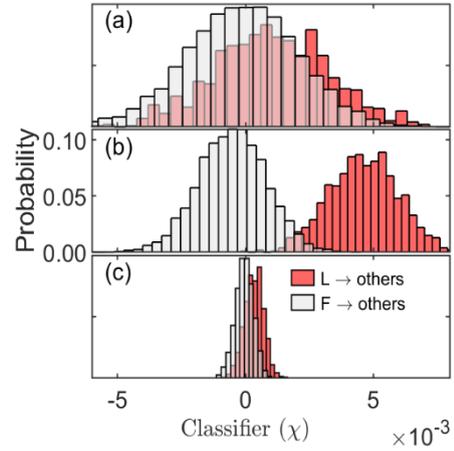

**Figure 3** Distribution of classifier $\chi$ for the leader and follower agents at three different noise levels $\eta_0$, (a) $0.1\pi$, (b) $1.2\pi$, and (c) $1.8\pi$

Time-delayed mutual information is capable of capturing both linear and nonlinear correlation between two time series and, hence, can be used as a measure of mutual coupling or information transmission between two processes [25]. However, it fails to consider shared history (and also common external driving effects) between two processes and may lead to spurious inferences of directed information transfer [42]. Let us consider the following two binary state time series $X$ and $Y$ whose initial states (at time $t = 1$) were chosen randomly, i.e.,

$$X(t=1) = x_1 = \begin{cases} 0 \text{ with probability } \frac{1}{2} \\ 1 \text{ with probability } \frac{1}{2} \end{cases},$$

and

$$Y(t=1) = y_1 = \begin{cases} 0 \text{ with probability } \frac{1}{2} \\ 1 \text{ with probability } \frac{1}{2} \end{cases},$$

where the process $X$ is autonomous, and its current state $x_t$ depends only on its own past state. Assume that the state of $X$ is switched from 0 to 1 and from 1 to 0 at each time step with probability of 100%. Conversely, the state of $Y$ at time $(t+1)$, $y_{t+1}$, has no direct relationship with its past state but depends on the state of $X$ at the preceding time step as $y_{t+1} = x_t$ with probability $\frac{1+c}{2}$, and $y_{t+1} = 1 - x_t$ with probability $\frac{1-c}{2}$ for some constant $-1 \leq c \leq 1$. Hence, easily one can find that $p(x_t = 0) = p(x_t = 1) = \frac{1}{2}$. After some manipulations, one can also find that $p(y_t = 0) = p(y_t = 1) = \frac{1}{2}$. Finally, joint probabilities $p(y_{t+1}, x_t)$ and $p(x_{t+1}, y_t)$ can also be computed (detail derivation can be found in Appendix).

Finally, using these marginal probabilities and joint probabilities in Eqs. (2a) and (2b) and setting $\tau = 1$, we get

$I(X(t); Y(t+1)) = I(X(t+1); Y(t)) = \frac{1}{2}[(1+c)\log_2(1+c) + (1-c)\log_2(1-c)]$.

Except $c = 0$, $I(X(t+1); Y(t)) > 0$. As the process $X$ is independent of $Y$ by definition, a positive time-delayed mutual information $I(X(t+1); Y(t))$, which indicates the information flow/transport [43] from $Y$ to $X$, contradicts with what we may expect. This means that time-delayed mutual information can misinterpret 'spurious transfer of information from $Y$ to $X$'. One should wonder about the source of this 'spurious information flow'. If we take a closer look at the model, we can realize that $y_t$ can tell the possible state (value) of $x_{t-1}$ because $y_t = x_{t-1}$ by definition, and furthermore $x_{t-1} = x_{t+1}$ by definition and hence $y_t$ can successively tell us everything about $x_{t+1}$, which creates the spurious information flow from $Y$ to $X$ [42,44]. One might criticize the two binary model itself discussed here as an extreme case but, in general, such spurious information flow can happen in systems where $X$ and $Y$ have shared history.

Hence to get the true influence of the past of $X$ on the present of $Y$, Schreiber [26] pointed out that we must condition on the past state(s) of $Y$ in Eq. (2a) using conditional mutual information. This leads to the definition of transfer entropy (TE), which overcomes this drawback of time-delayed mutual information. TE from the random variable $Y$ to $X$ is defined for a time lag $\tau$ [38]:

$$\text{TE}_{X \to Y} = I(y_{t+\tau}; x_t | y_t)$$
$$= \sum_{y_{t+\tau}, x_t, y_t} p(y_{t+\tau}, x_t, y_t) \log_2 \frac{p(y_{t+\tau} | y_t, x_t)}{p(y_{t+\tau} | y_t)}, \quad (3)$$

where $I(.;.|.)$ and $p(.|.)$ represent conditional mutual information and conditional probability, respectively. TE from $X$ to $Y$ quantifies how well the random variable $X$ can predict the outcome of the variable $Y$ using the past of both $Y$ and $X$, rather than using the past of $Y$ alone.

Going back to the binary example, one may easily compute the joint probabilities $p(y_{t+1}, y_t, x_t), p(y_{t+1}, y_t)$ which are necessary to compute $\text{TE}_{X \to Y}$ (see Appendix for detail calculations).

Setting $\tau = 1$ in Eq. (3), we then get

$\text{TE}_{X \to Y} = \frac{1}{2}c[(1+c)\log_2(1+c) - (1-c)\log_2(1-c)]$
$\qquad - \frac{1}{2}(1+c^2)\log_2(1+c^2) > 0$

except $c = 0, \pm 1$. Note that $x_t$ can tell the state of $x_{t+1}$ perfectly, resulting in $p(x_{t+1} | x_t) = 1$. This implies that $x_{t+1}$ contains no uncertainty to be reduced once $x_t$ is known, hence $H(x_{t+1} | x_t) = H(x_{t+1} | x_t, y_t) = 0$. These yields $\text{TE}_{Y \to X} = 0$. The causal structure described in the example does not allow information to flow from $Y$ to $X$, which is correctly identified by TE as $\text{TE}_{Y \to X} = 0$. Also, for $c \neq 0$, $\text{TE}_{X \to Y} > 0$ which is along what we expect [44].

A positive $\text{TE}_{X \to Y}$ indicates that past of $X$ contains some useful information about $y_{t+\tau}$ which is not contained in the past of $Y$, which indicates the causal influence of $X$ on $Y$ [26]. Since a follower agent follows the motion of a leader but the converse is not true, hence a leader can predict the motion of a follower more precisely. But a follower cannot predict a leader with such precision. Also, net TE from $X$ to $Y$ is defined as $\text{NTE}_{X \to Y} = \text{TE}_{X \to Y} - \text{TE}_{Y \to X}$. Hence a positive $\text{NTE}_{X \to Y}$ indicates that $Y$ follows $X$. In other words, one may classify $X$ as a leader and $Y$ as a follower when $\text{NTE}_{X \to Y} > 0$.

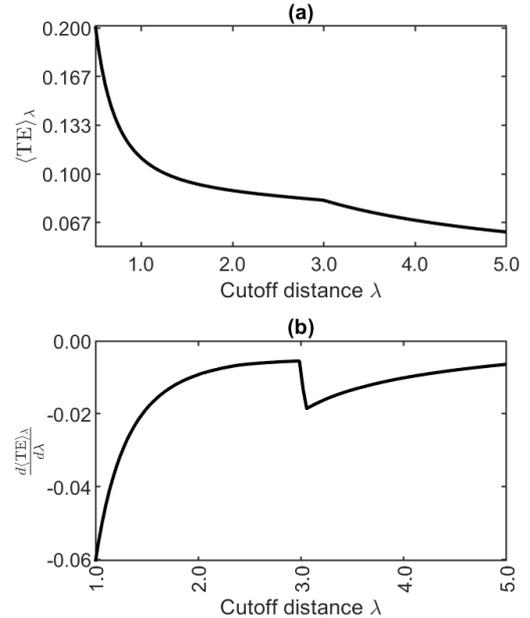

Figure 4 (a) Averaged TE $< \text{TE} >_\lambda$ as a function of cutoff distance $\lambda$ for $R = 3$ and $\eta_0 = 1.2\pi$. (b) $\frac{d<\text{TE}>_\lambda}{d\lambda}$ as a function of cutoff distance $\lambda$ for $R = 3$ and $\eta_0 = 1.2\pi$. It has been found that $\frac{d<\text{TE}>_\lambda}{d\lambda}$ exhibits a minimum near the actual interaction radius $R$.

**Models:**

Vicsek et al. proposed a flocking model displaying a transition to collective motion [11]. The system consists of a two-dimensional square box of length $L$ with periodic boundary conditions, and $N$ self-propelled agents move within the box with the same constant speed $v$, and at time $t = 0$ the agents are positioned and oriented randomly [45-47]. In addition, the velocities $\{\vec{v}_i\}$ of the agents are updated simultaneously with time increment $\Delta t$. The position of the agent $i$ ($i = 1, 2, \ldots, N$) is updated according to

$$\vec{r}_i(t+1) = \vec{r}_i(t) + \vec{v}_i(t)\Delta t, \qquad (3a)$$

where $\vec{r}_i(t)$ and $\vec{v}_i(t)$ denote the position and velocity of the agent $i$ at time $t$, respectively. The velocity $\vec{v}_i(t+1)$ of the agent $i$ at time $(t+1)$ is calculated to have the constant value $v$ and an orientation $\theta_i(t+1)$ given by

$$\theta_i(t+1) = <\theta(t)>_{R,\vec{r}_i(t)} + \Delta\theta_i(t), \qquad (3b)$$

where $<\theta(t)>_{R,\vec{r}_i(t)}$ represents the averaged orientation over agents (including the agent $i$ itself) which are located within a circle of radius $R$ at time $t$ centered on $\vec{r}_i(t)$ [Fig. 1]. The motion of the agent is perturbed by the noise term $\Delta\theta_i(t)$ which is a random number uniformly distributed in the range $\left[-\frac{\eta_0}{2}, \frac{\eta_0}{2}\right]$, where $\eta_0$ is considered as a temperature-like parameter.

To study the phase transition between ordered states of motion at low levels of noise and disordered motion at high level of noise, a natural order parameter can be the absolute value of the normalized mean velocity ($\phi$), i.e., $\phi = \frac{1}{Nv}\left|\sum_{i=1}^{N}\vec{v}_i\right|$. In the disordered phase the value of $\phi$ is close to zero and it is close to unity in the ordered phase [46]. In the VM, all the agents have the same effect on their neighbors, which means there exists no leader-follower relationship between agents.

To create a mathematical model of the leader-follower relationship between moving agents, one first needs to define what leadership means. Garland et al. [48] has pointed out three categories of leadership, that is, (i) structural leadership, (ii) informed leadership, and (iii) emergent leadership.

*Structural leadership* relies on the structure of the interaction between agents in a group. The existence of structural leadership implies that the rules of interaction are encoded with information about which agent(s) is (are) the leader. Age, gender, size, reproductive state, lifespan etc., could be the decisive factors for this kind of leadership. For instance, a queen in a colony of honeybee has ten times lifespan relative to workers and is the only agent that can lay fertilized eggs [49,50]. An elephant clan is guided by a matriarch [51]. The leader in a baboon group is occupied by a big male [52].

*Informed leadership* is the case where a member of the group has additional information that the other members do not. By incorporating this information into its actions, the collective will be skewed to the leader's direction. For example, animals migrate to find food, nestle, or find a more favorable living or breeding condition [53-57].

During this migration, an individual or a subset of the migrating group may have prior knowledge about the migration routes, source of water, or locations of predators [58]. Taking advantage of this prior knowledge, that small part of the migrating group influences the movement of the whole group by changing speed, the direction of movement, or through other signaling indications [48,59,60]. That individual or small group can be considered as the leader(s) of the group, and such kind of leadership is termed as the informed leadership.

Even in the absence of the social structure or differential information, another form of leadership may arise, termed as *emergent leadership*. When the motion of the agents is influenced by the individuals that are in front of them, for example, the frontal agents may be more influential even though they have no additional information, motivation, or status. In the VM [11] or the modified VM studied in [9,16], the interaction domain is assumed to be a circle (for 2-D motion) [Fig. 1] of a specified radius and the motion of an agent is influenced by its surroundings by assuming a 360° angle of view. But all species of animals do not have a 360° view of their surroundings [61]. For example, the field of view of grey-headed albatross is about 270° [62], and *Dasyatis sabina* fish is about 327° [63]. Hence instead of a whole circular interaction domain, Durve et al. [61] has considered a section of the circle (field of view) as the interaction domain, and this section moves with the change of heading of the agent. In other words, the motion of an agent can be influenced by the motion of another agent if the latter one lies within a certain distance and field of view of the 1st agent. Hence the latter agent should have an asymmetric influence on the motion of its posterior agents that constitutes the emergent leadership. Yet there is no social rule signifying that this agent should be a leader (i.e. it may have occurred by chance), and that agent has no additional navigational knowledge compared to other agents.

The modified Vicsek model studied by Basak et al. [9] is an example of structural leadership where the sociality matrix $\boldsymbol{w}$ was introduced to emulate asymmetric interaction between agents. The orientation angle of the agent $i$ was updated at each time step:

$$\theta_i(t+1) = <\theta(t-\kappa+1)>_{R,\boldsymbol{w},\vec{r}_i(t)} + \Delta\theta_i(t), \qquad (4)$$

where $<\theta(t-\kappa+1)>_{R,\boldsymbol{w},\vec{r}_i(t)}$ represents the weighted alignment of agents averaged over other agents which are located within the circle of radius $R$ at time $(t-\kappa+1)$ centered at $\vec{r}_i(t)$. Naturally, there should be a finite time difference between the interaction time and timescale of movement of agents which was characterized by the variable $\kappa$ in Eq. (4). The element $w_{ij}$ of the sociality matrix $\boldsymbol{w}$ corresponds to the interaction strength that the agent $i$ exhibits on $j$. If $i$ is a leader and $j$ is a follower, then $w_{ij} > w_{ji}$ which characterizes the dominance of leader agents. The interaction strength between two follower agents, two leader agents, and also from follower to leader agents were set to 1, i.e., $w_{FF} = w_{LL} = w_{FL} = 1$. Also, $w_{LF}$ was set to 1.05 for the results reviewed in this paper. The interactions among the agents were set to occur only when the agents are within the interaction radius $R$.

**Identifying leader agents**

Hypothetically in a group of mutually interacting agents, the group decision-making should be strongly influenced by the leader(s) of the group [64]. Hence the agents that on average lead other group members could be identified as leaders of the group. A directed network scheme has been introduced in [16] to infer leader-follower relationships where each node corresponds to an agent and the weighted directed edges represent the role of agents (leader or follower) in the pair. Based on the interaction strength, the weighted adjacency matrix $\boldsymbol{W}$ is formulated. For a pair of agents $i$ and $j$, let $I_{ij}$ represent the influence (measured using cross correlation (CC), TE, time delayed mutual information (TDMI) etc.,) that the agent $i$ exhibits on $j$. Then the element $W_{ij}$ of the matrix $\boldsymbol{W}$ is chosen in such a way that $W_{ij} = I_{ij}$ if $i$ is detected as the leader and $j$ as the follower ($I_{ij} > I_{ji}$), otherwise $W_{ij} = 0$. Hence both $W_{ij}$ and $W_{ji}$ cannot be positive simultaneously but their values could be zero if neither of the two is detected as a leader. Then the average pairwise interaction for the agent $i$ is defined as:

$$\chi^i = \frac{1}{N-1} \sum_{j=1}^{N} [W_{ij} - W_{ji}],$$

where $N$ represents the number of agents in the group. This average pairwise asymmetric interaction for the agent $i$ would be negative if the agent is dominated on average by other(s), and positive if it dominates other(s) on average. $\chi^i$ is then used as the classifier. Setting a threshold $\epsilon$ on the value of $\chi^i$, for example $\epsilon = 0$, agents are assigned as leaders (followers) whenever $\chi^i > \epsilon$ ($\chi^i \leq \epsilon$).

In the weighted adjacency matrix $\boldsymbol{W}$, the element $W_{ij}$ is set to zero if the scheme identifies $i$ as the follower and $j$ as the leader. That means the effect of a follower on a leader is disregarded, and hence some information is lost which may have an effect on the performance of leader-follower classification.

A much simpler classification algorithm has been presented in [9] where the identification of an agent is based on the following classifier:

$$\chi^i = \frac{1}{N-1} \sum_{j(\neq i)} [I_{ij} - I_{ji}],$$

where $N$ represents the number of agents and $I_{ij}$ represents TE (or CC) from the agent $i$ to $j$. The basis of the leader-follower relationship is that the influence of a leader is much greater on its followers compared to that of a follower on the leader. Hence the net quantity ($I_{ij} - I_{ji}$) is a good candidate for classifying leader and follower agents. The agents having higher $\chi$ values are the candidates for the leaders of the group.

The value of $\chi^i$ for each agent is compared to a threshold value $\epsilon$ to determine their identities. An agent for which $\chi^i$ is higher than the threshold $\epsilon$ is identified as a leader, otherwise as a follower. This classification result is then compared to the ground truth to obtain the number of true positives and number of false positives for the chosen $\epsilon$. The exact theoretical value of $\epsilon$ is positive whenever $i$ is a leader of $j$, negative when $j$ is a leader of $i$, and zero when there is no leader-follower relationship between $i$ and $j$. Therefore, a natural value to choose for $\epsilon$ is 0. However, one can tune $\epsilon$ when it is desired to do so. A larger $\epsilon$ could be desirable, for example, if false negatives are preferred over false positives. To show the classification performance of a classifier over all possible $\epsilon$, a receiver-operating characteristic (ROC) curve is used. It is obtained by plotting the true-positive rate (TPR) versus false-positive rate (FPR) at different values of $\epsilon$, where TPR and FPR are defined as follows [65]:

$$\text{TPR} = \frac{\text{True positive}}{\text{True positive} + \text{False negative}},$$

$$\text{FPR} = \frac{\text{False positive}}{\text{False positive} + \text{True negative}}.$$

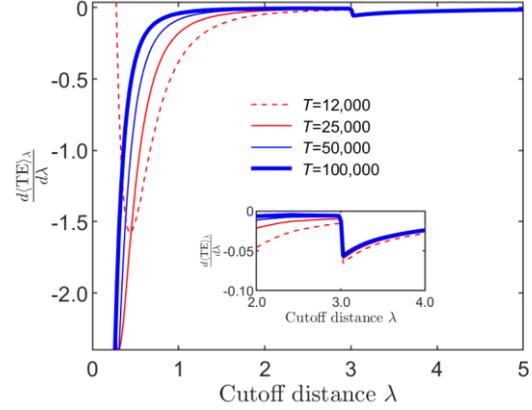

**Figure 5** $\frac{d\langle TE\rangle_\lambda}{d\lambda}$ as a function of cutoff distance $\lambda$ for different $T$ at $\eta_0 = 1.2\,\pi$ and $R = 3$. For shorter $T$ a global minimum of $\frac{d\langle TE\rangle_\lambda}{d\lambda}$ has been observed at short $\lambda$. Hence global minimum scheme identifies inaccurate interaction radius when $T$ is short.

A good classifier has the highest TPR along with the lowest FPR. Hence, the closer the ROC curve gets to the top-left corner, the better the classifier is [Fig. 2]. To quantify the diagnostic ability of a binary classifier and also to compare the performance of different binary classifiers one can use area under ROC curve (AUC). An AUC score of 0.5 represents the performance of a random classifier (which means that the classifier has no class separation capability at all) and the maximum value of AUC (1.0) corresponds to a perfect classifier [66].

**Some studies for leader-follower classifications based on trajectories of agents**

Using synthetic trajectory data of zebrafish pairs swimming in 2D, net transfer entropy was found to be the most accurate classifier for leader-follower relationship, compared to cross-correlation and extreme-event synchronization [67]. For two different cases, that is, a fixed box size while the number of agents is varied, and a fixed density while the number of agents is varied, it was shown using a modified VM that the classification score decreases as the number of agents increases. This is due to the fact that, for any pair of agents for which TE is estimated, the motion of an agent is influenced by other agents [9]. Interestingly at very low, and very high levels of noise AUC scores are close to 0.5 (i.e., same as coin toss). The reasoning was, at low noise levels, the probability of encountering of two agents depends largely on the initial position of the agents. Also, the leader and follower agents get aligned in the same direction quickly once they encounter each other, resulting in the same symbols in their time series. Hence TE between agents depends on the initial configuration of the agents to a great extent resulting in higher variances in the distributions of leaders and followers, making the classification difficult [Fig. 3a]. On the contrary, at very high noise levels, the distributions become indistinguishable with much smaller variances that produce a low AUC score [Fig. 3c].

Mwaffo et al. [11] revised the VM to discuss informed leadership where the motion of leader agents does not depend on the other group members, and instead moves towards a specific direction with the presence of noise. The headings of the follower agents were updated based on the response of the group in presence of noise.

But leadership may change over time from one agent to another i.e., a leader agent may emerge as a follower after leading the group for a certain period of time, and a new follower takes the charge of the group as the leader. Butail et al. [68] have studied the switching leadership in the collectives where the motion (orientation) of leader agent is independent of the influence of other agents whereas the orientation of a follower agent is calculated based on its instantaneous neighboring agents. The role of a leader and a follower agent could be switched randomly over a simulation. It has been shown that TE is capable of partitioning of time series data to detect leadership switches in collective behaviors.

Recent image sensing technologies such as image processing and global positioning system (GPS) have enabled us to collect various kinds of data of animal groups such as bats [69,70]. Studying the 3D trajectories of wild bats flying in pairs, higher transfer entropy from the bat flying in front to the bat flying in the rear was confirmed, which provides the evidence that the relative spatial positioning is the key in navigational leadership [70].

However, in all above-mentioned studies, any distance evaluating transfer entropy between two agents was not considered. Recently, we showed [9] using a modified VM that the classification scores of leaders and followers increase significantly by combining the identified interaction domain in the TE estimation compared to the conventional scheme where the distance information is not considered in the estimation.

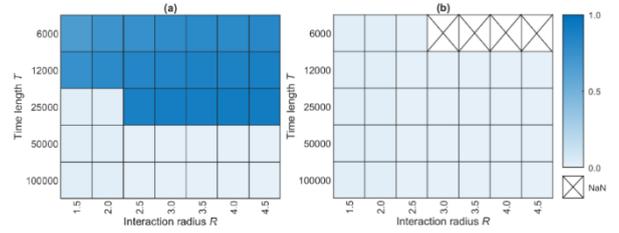

**Figure 6** Relative error landscape as a function of interaction radius $R$ and data length $T$ using (a) global minimum scheme, and (b) convexity score scheme at $\eta_0 = 1.2\,\pi$ with $\delta = 1 \times 10^{-4}$ and $M = \{M | 2 \leq M \leq 30\}$. Global minimum scheme has high relative error compared to convexity score scheme for shorter data length. Cross-marked boxes 'NaN' mean that the scheme fails to identify the interaction radius.

We proposed information-theoretic schemes to infer the domain of interaction using the trajectories of the agents. The schemes are based on a quantity termed as the 'cutoff distance $\lambda$', which is defined as a predefined maximum distance up to which the interaction between agents is considered for the estimation of TE [9,71]. In fact, for a pre-defined cutoff distance $\lambda$, the TE between two agents is computed as follows: only if the distance between two agents at time $t$ is less than or equal to $\lambda$, their time series of the two agents are used to estimate probability distributions at that time instance [9]. Then the value of $\lambda$ is varied and TE between agents is computed as a function of $\lambda$. In the problem setting, the domain of interaction was considered as a circle of radius $R$, which is typically unknown.

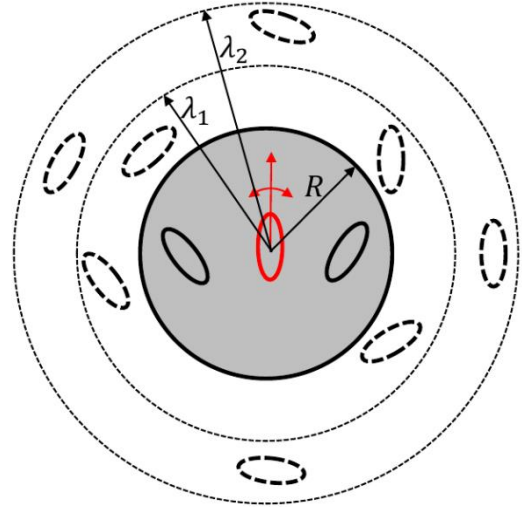

**Figure 7** This schematic diagram of how the number of interacting (black solid ovals)/ non-interacting agents (dotted ovals) depends on cutoff distance. The shaded region represents the interaction zone of the red-colored agent. For simplicity, it is assumed that the interaction strength inside the interaction zone remains the same and no interaction exists outside the shaded zone. As the cutoff distance $\lambda$ increases over interaction radius $R$, the number of non-interacting agents increases.

For very long data set ($T = 100{,}000$) it was shown that the average TE as a function of $\lambda$ over all pairs of agents, $<\text{TE}>_\lambda$, drops for a small interval of $\lambda$ when $\lambda$ is small. Then $<\text{TE}>_\lambda$ is almost flat up to $\lambda \cong R$, and for the region of $\lambda > R$, $<\text{TE}>_\lambda$ decreases again [9]. It was also found that the derivative of $<\text{TE}>_\lambda$ with respect to $\lambda$ i.e., $\frac{d<\text{TE}>_\lambda}{d\lambda}$ exhibits a minimum near $\lambda = R$ [Fig. 4]. Here the proposed scheme is: the estimated interaction radius $\hat{R}$ is defined as the corresponding $\lambda$ value for which $\frac{d<\text{TE}>_\lambda}{d\lambda}$ exhibits a local minimum i.e.,

$$\hat{R} = \text{argmin}_\lambda \frac{d<\text{TE}>_\lambda}{d\lambda}, \text{ with } \frac{d^2<\text{TE}>_\lambda}{d\lambda^2} = 0.$$

For a VM system of ten agents where a single agent serves as a leader and rest are the followers, it was found that the above-mentioned protocol successfully identifies the interaction radius [9]. In data acquired from experiments, the data acquisition process is discrete, and the observed time scale does not exactly match the timescale of interaction in general. Hence one may not be able to sample the data at every relevant time step. Consider the case where one can sample the data once every $\tau_{obs}$ time steps. When the time scale of system dynamics in Eq. (4) and the observation time scale $\tau_{obs}$ are the same, cross correlation (CC) can also estimate the underlying interaction radius as TE does except at low noise and short interaction radius $R$. However, for low and moderate noise levels, it was observed that the performance of CC gradually decreases as the observation timescale is higher than the interaction timescale. But when the noise level gets much higher, both the schemes fail to detect the underlying interaction domain for the case of the observation time $\tau_{obs} > 1$ [9].

This scheme in inferring interaction domain can be used for a system containing a large number of agents, and even for a system where there exists no leader-follower relationship among the agents [9].

The global minimum of $\frac{d<\text{TE}>_\lambda}{d\lambda}$ scheme is capable of identifying actual interaction radius for sufficiently long trajectory data ($\geq 50{,}000$). However, it was found that for shorter length ($T = 12{,}000 - 25{,}000$), $\frac{d<\text{TE}>_\lambda}{d\lambda}$ fails to capture the correct interaction domain with the global minimum at shorter $\lambda$ than the underlying desired value [Fig. 5]. Since in actual experiments it is sometimes not possible to get a set of long trajectories, we thus developed a different scheme based on the convexity score of points at coarse-grained level [71]. The convexity score scheme was found be robust against fluctuations of $\frac{d<\text{TE}>_\lambda}{d\lambda}$. This scheme is based on the fact that $\frac{d<\text{TE}>_\lambda}{d\lambda}$ as a function of $\lambda$ is convex near the actual interaction radius $R$.

The convexity score $\kappa(\lambda_i)$ based on the $M$ neighboring points of $\lambda_i$ was defined as follows:
$\kappa(\lambda_i) = \frac{1}{M}\sum_{m=1}^{M}\sigma_i(m)$ where
$\sigma_i(m) =$

$$\begin{cases} +1, \text{if } f(\lambda_{i-m}) - f(\lambda_i) > \delta \text{ and } f(\lambda_{i+m}) - f(\lambda_i) > \delta \\ -1, \text{if } (\lambda_i) - f(\lambda_{i-m}) > \delta \text{ and } (\lambda_i) - f(\lambda_{i+m}) > \delta \\ \phantom{+}0, \text{ otherwise} \end{cases}$$

Here $\delta$ represents the non-negative small number, $f(\lambda) = \frac{d<\text{TE}>_\lambda}{d\lambda}$, and $-1 \leq \kappa(\lambda_i) \leq 1$. Finally, the point $\lambda = \lambda_i$ around which the convexity score $\kappa(\lambda_i)$ is the maximum was chosen as the estimated interaction radius $\hat{R}$. The optimal values of the parameter $M$ were chosen in such a way that the cost function $C(M) = \sum_T \sum_{T'} |\hat{R}(M,T) - \hat{R}(M,T')|$ is the minimum. It should be noted that, those $T$ for which $\hat{R}(M,T)$ was not chosen uniquely due to the degeneracy of $\kappa$, or $\hat{R}(M,T)$ was undefined due to the absence of strongly convex part of $\frac{d<\text{TE}>_\lambda}{d\lambda}$, were excluded from the computation of optimal $M$ [71].

To compare the performances of global minimum scheme and convexity score scheme, relative error has been used. Relative error ($\Delta R$) is defined as follows [9]

$$\Delta R = \frac{|R - \hat{R}|}{R},$$

where $R$ is the actual interaction radius and $\hat{R}$ is the identified interaction radius. It was shown using a relative error diagram that the convexity score scheme identifies the interaction radius satisfactorily, (at moderate noise levels) even when data length is short for those data lengths $T$ where the global minimum scheme has high relative error [Fig. 6] [71]. Also, when the data length $T$ is very short ($\approx 6{,}000$), the convexity score fails to identify the interaction radius for large $R$ [Fig. 6(b)].

Hence, a measure of convexity of $\frac{d<\text{TE}>_\lambda}{d\lambda}$ is expected to be useful in identifying interaction radius. In brief, for the sake of simplicity, let's assume that the interaction strength remains the same inside the interaction zone (shaded region in [Fig. 7]) and no interaction exists outside it. Then, the analytical expression of average TE, $<\text{TE}>_\lambda$ as a function of $\lambda$ is easily acquired, showing $<\text{TE}>_\lambda$ remains the same for $\lambda \leq R$ that yields $\frac{d<\text{TE}>_\lambda}{d\lambda} = 0$ [71]. In VM, as seen in Fig. 4a, $<\text{TE}>_\lambda$ drops rapidly for $\lambda \ll R$ with respect to $\lambda$ and turns to be flat, yielding $\frac{d<\text{TE}>_\lambda}{d\lambda} \simeq 0$, near $\lambda \cong R$. As $\lambda$ increases more than $R$, the number of non-interacting agents that behave independently increases. As independent agents do not share information, hence $<\text{TE}>_\lambda$ decreases. As a result, $\frac{d<\text{TE}>_\lambda}{d\lambda}$ becomes negative. Eventually $<\text{TE}>_\lambda$ converges to zero as $\lambda \to \infty$. Consequently, $\frac{d<\text{TE}>_\lambda}{d\lambda}$ approaches to zero. Hence, a kink/minimum is expected at $\lambda = R$ in the $\frac{d<\text{TE}>_\lambda}{d\lambda}$ curve.

**Conclusion and perspectives:**
In recent studies it has been elucidated that information-theoretic measures are successful in identifying leaders and followers [9,16,33,34,67,70-72]. However, these measures have disadvantages in the cases where there are multiple interacting variables or when the amount of available data is limited. Knowledge of the interaction domain greatly increases the accuracy of information-theoretic measures for classifying leaders and followers by incorporating only the relevant portions of time series in computing the measures [9]. This interaction domain can be inferred using transfer entropy with cutoff function, however, information-theoretic measures themselves have some drawbacks that still need to be addressed, both for inferring interaction domain and for inferring leadership. For example, information-theoretic measures such as transfer entropy require discretization of the data, and none of the studies in this area of research have addressed the question of computing the optimal discretization of data for computing transfer entropy. Furthermore, the transfer entropy computation requires a parameter $\tau$ to understand the time scale of delay between the leader's action and the follower's response. Without knowledge of a suitable $\tau$, it is not possible to infer leadership or interaction domain precisely using transfer entropy. In addition, one assumption of transfer entropy computation is stationarity, which is not necessarily guaranteed in some systems.

Another drawback of transfer entropy is that it cannot necessarily quantify the "flow of information". Using a simple binary model, it was demonstrated that the transfer entropy between two processes is not localized in the way that the expression of "transfer from one to another" [73], hence transfer entropy is not appropriate to characterize 'information flow' in some cases. However, in recent works [17,74], it is shown that transfer entropy can be decomposed into two distinct modes of information flow, namely intrinsic and synergistic. This intrinsic information from X to Y is designed to extract solely the information flow from X to Y, buried in transfer entropy. Hence, this intrinsic information is more fundamental and appropriate to describe information flow between two processes.

One example of a system which exhibits above-mentioned challenges is the aggregation of dictyostelium discoideum cells. Multiple variables characterize the cell activity such as its direction of motion, its speed, and its response to the chemotractant signal cAMP, and its release of additional cAMP [75-77]. A single quantity e.g., direction of motion may not be enough to capture the entire information flow in such a system. Furthermore, the activity of dictyostelium discoideum cells in response to cAMP changes over time. Lastly as real systems, it is not clear what is a suitable $\tau$ or symbolization, and the ability of the data to characterize information flow depends highly on the time-resolution of the experiment and the variables chosen for computing information flow. A fully systematic approach to interring leadership from experimental data should address the above issues and in general systems of collectively moving agents.

**Acknowledgements**
This work was supported by a Grant-in-Aid for Scientific Research on Innovative Areas 'Singularity Biology (No. 8007)' (Grant No. 18H05413), MEXT, the research program of 'Five star Alliance' in 'NJRC Matter and Dev' (Grant No. 20191062-01), the JSPS (Grant Nos. 25287105 and 25650044, T.K.), and the JST/CREST (Grant No. JPMJCR1662, T.K.)

**Conflicts of Interest**
All authors declare that they have no conflict of interest.

**Author Contributions**
T.K., S.S. designed the project over discussions with K.H. who provided a series of movies of cAMP dynamics of dictyostelium discoideum cells. K.H. contributed to the design of the research objectives. U.S.B. and S.S. conducted analyses and experiments. M.H. analyzed causal relation of the image data. U.S.B. created all figures. U.S.B., S.S. and T. K. wrote the manuscript. All authors confirmed the contents of the manuscript.

**Appendix**
From the model it is obvious that $p(x_t = 1) = p(x_t = 0) = \frac{1}{2}$. Now,

$$\begin{aligned} p(y_t = 1) &= \sum_{x_t} p(y_t = 1, x_t) \\ &= \sum_{x_t} p(y_t = 1|x_t) p(x_t) \\ &= p(y_t = 1|x_t = 1) p(x_t = 1) \\ &\quad + p(y_t = 1|x_t = 0) p(x_t = 0) \\ &= p(y_t = 1|x_{t-1} = 0) p(x_t = 1) \\ &\quad + p(y_t = 1|x_{t-1} = 1) p(x_t = 0) \\ &= \frac{1+c}{2} \times \frac{1}{2} + \frac{1-c}{2} \times \frac{1}{2} \\ &= \frac{1}{2}. \end{aligned}$$

Similarly, one can easily find $p(y_t = 0) = \frac{1}{2}$. Now let us compute the joint probability $p(y_{t+1}, x_t)$. Here,

$$\begin{aligned} p(y_{t+1} = 1, x_t = 1) &= p(y_{t+1} = 1|x_t = 1) p(x_t = 1) \\ &= \frac{1+c}{2} \times \frac{1}{2} \\ &= \frac{1+c}{4} \\ &= p(y_{t+1} = 0, x_t = 0) \end{aligned}$$

and, likewise

$$p(y_{t+1} = 1, x_t = 0) = p(y_{t+1} = 0, x_t = 1) = \frac{1-c}{4}.$$

Hence,

$$\begin{aligned} \text{TDMI}(X \to Y) &= \text{MI}(y_{t+1}, x_t) \\ &= \sum_{y_{t+1}} \sum_{x_t} p(y_{t+1}, x_t) \log_2 \frac{p(y_{t+1}, x_t)}{p(y_{t+1}) p(x_t)} \\ &= \frac{1}{2} [(1+c) \log_2(1+c) \\ &\quad + (1-c) \log_2(1-c)]. \end{aligned}$$

One can also straightforwardly compute TDMI($Y \to X$):

$$\begin{aligned}
\text{TDMI}(Y \to X) &= \text{MI}(x_{t+1}, y_t) \\
&= \sum_{x_{t+1}} \sum_{y_t} p(x_{t+1}, y_t) \log_2 \frac{p(x_{t+1}, y_t)}{p(x_{t+1})p(y_t)} \\
&= \frac{1}{2}[(1+c)\log_2(1+c) \\
&\quad + (1-c)\log_2(1-c)].
\end{aligned}$$

To compute $\text{TE}_{X \to Y}$, first we need to determine the joint probabilities $p(y_{t+1}, y_t, x_t)$. Here,

$p(y_{t+1}, y_t, x_t) = p(y_{t+1}, y_t | x_t) p(x_t) = p(y_{t+1}|x_t) p(y_t|x_t) p(x_t)$, as $y_{t+1}$ and $y_t$ are conditionally independent given $x_t$. Hence,

$$\begin{aligned}
p(y_{t+1} &= 1, y_t = 1, x_t = 1) \\
&= p(y_{t+1} = 1 | x_t = 1) p(y_t = 1 | x_t = 1) p(x_t = 1) \\
&= p(y_{t+1} = 1 | x_t = 1) p(y_t = 1 | x_{t-1} = 0) p(x_t = 1) \\
&= \frac{1+c}{2} \times \frac{1-c}{2} \times \frac{1}{2} \\
&= \frac{1-c^2}{8}.
\end{aligned}$$

Likewise, one can find that,

$$\begin{aligned}
p(y_{t+1} = 1, y_t = 1, x_t = 0) &\\
= p(y_{t+1} = 0, y_t = 0, x_t = 1) &\\
= p(y_{t+1} = 0, y_t = 0, x_t = 0) &\\
= \frac{1-c^2}{8}&, \\
p(y_{t+1} = 1, y_t = 0, x_t = 1) &\\
= p(y_{t+1} = 0, y_t = 1, x_t = 0) &\\
= \frac{(1+c)^2}{8}&, \\
p(y_{t+1} = 0, y_t = 1, x_t = 1) &\\
= p(y_{t+1} = 1, y_t = 0, x_t = 0) &\\
= \frac{(1-c)^2}{8}&.
\end{aligned}$$

Again,

$$p(y_{t+1}, y_t) = \sum_{x_t} p(y_{t+1}, y_t, x_t)$$

Hence, $p(y_{t+1} = 1, y_t = 1) = p(y_{t+1} = 1, y_t = 1, x_t = 1) + p(y_{t+1} = 1, y_t = 1, x_t = 0) = \frac{1-c^2}{8} + \frac{1-c^2}{8} = \frac{1-c^2}{4}$. Likewise,

$$p(y_{t+1} = 1, y_t = 0) = p(y_{t+1} = 0, y_t = 1) = \frac{1+c^2}{4},$$
$$p(y_{t+1} = 0, y_t = 0) = \frac{1-c^2}{4}.$$

Finally, we can get

$$\begin{aligned}
\text{TE}_{X \to Y} &= \sum_{y_{t+1}} \sum_{y_t} \sum_{x_t} p(y_{t+1}, y_t, x_t) \log_2 \frac{p(y_{t+1}, y_t, x_t) p(y_t)}{p(y_{t+1}, y_t) p(y_t, x_t)} \\
&= \frac{1}{2} c[(1+c)\log_2(1+c) - (1-c)\log_2(1-c)] \\
&\quad - \frac{1}{2}(1+c^2)\log_2(1+c^2).
\end{aligned}$$